\documentclass[twocolumn,preprintnumbers,aps]{revtex4}
\usepackage{graphicx}

\newcommand{\bi}{\bibitem}
\newcommand{\be}{\begin{eqnarray}}
\newcommand{\ee}{\end{eqnarray}}

\begin{document}

\title{Primordial black holes and the observed Galactic 511 keV line}

\author{Cosimo Bambi$^{\rm 1}$}
\author{Alexander D.~Dolgov$^{\rm 2,3,4}$}
\author{Alexey A.~Petrov$^{\rm 1,5}$}

\affiliation{$^{\rm 1}$Department of Physics and Astronomy, 
Wayne State University, Detroit, MI 48201, USA\\
$^{\rm 2}$Istituto Nazionale di Fisica Nucleare, 
Sezione di Ferrara, I-44100 Ferrara, Italy \\
$^{\rm 3}$Dipartimento di Fisica, 
Universit\`a degli Studi di Ferrara, I-44100 Ferrara, Italy \\
$^{\rm 4}$Institute of Theoretical and Experimental Physics,
113259 Moscow, Russia \\
$^{\rm 5}$Michigan Center for Theoretical Physics, University of Michigan,
Ann Arbor, MI 48109, USA}

\date{\today}

\preprint{WSU-HEP-0801}

\begin{abstract}
The observed 511 keV line from the Galactic Bulge is a real challenge for 
theoretical astrophysics: despite a lot of suggested mechanisms, there is 
still no convincing explanation and the origin of the annihilated positrons 
remains unknown. Here we discuss the possibility that a population of slowly 
evaporating primordial black holes with the mass around 
$10^{16}-10^{17}$~g ejects (among other particles) low--energy positrons 
into the Galaxy. In addition to positrons, we have also calculated the 
spectrum and number density of photons and neutrinos produced by such black 
holes and found that the photons are potentially observable in the near 
future, while the neutrino flux is too weak and below the terrestrial and 
extra--terrestrial backgrounds. Depending on their mass distribution, such 
black holes could make a small fraction or the whole cosmological dark matter. 
\end{abstract}

\maketitle

{\sc Introduction ---} It is now clear that in the central
region of the Galaxy electron--positron annihilation proceeds
at a surprisingly high rate. Confirming precedent 
measurements~\footnote{The first evidences of the Galactic 511 
keV line date back to the '70s~\cite{old-1}. Over the past 40
years, there have been numerous publications on the observation
of this line, see e.g.~\cite{old-2} and references therein for 
more recent measurements.}, the SPI spectrometer on the INTEGRAL 
satellite has detected an intense 511 keV gamma ray line flux 
(Bulge component)~\cite{spi-1, spi-2}
\be\label{flux}
\Phi_{511 \; {\rm keV}} = 1.07 \pm 0.03 \cdot 10^{-3} \; 
{\rm photons \; cm^{-2} \, s^{-1}}
\ee
with a width of about 3 keV, consistent with two dimensional
gaussian distribution aligned with the Galactic Center and with
a full width at half maximum (FWHM) of about 8$^\circ$. More recently, 
the SPI spectrometer has also provided evidence for the disk or halo 
component~\cite{spi-halo} of this line.

Non--relativistic positrons in the interstellar medium can either 
directly annihilate with electrons or form positronium. 
In the first case they produce two 511 keV photons,
while in the second case the annihilation channels are different
for para--positronium (formed with 25\% probability) and ortho--positronium
(formed with 75\% probability). Para--positronium decays into 
two 511 keV $\gamma$s, while ortho--positronium decays into three photons
with continuous spectrum. From the relative intensities 
of the 511 keV line and the three $\gamma$ continuum, one can
deduce that the total annihilation rate is 3.6 times larger than 
the one arising from consideration of flux~(\ref{flux}) 
only~\cite{beacom}. Thus, assuming that the Solar System lies 
at a distance $r = 8.5$~kpc from the Galactic Center, we can 
conclude that the total rate is about $3 \cdot 10^{43}$ 
annihilations per second.

From the theoretical point of view, the issue is to identify
the source of these galactic positrons. Many production mechanisms
have been suggested, but so far no one is completely satisfactory.
Among conventional mechanisms, which do not demand new physics, 
there are type Ia supernovae~\cite{Milne:1999ri}, low mass X--ray 
binary systems~\cite{spi-1} (see also~\cite{Guessoum:2006fs})
and the energetic electrons and photons created by accretion on the
super--massive black hole at the Galactic Center~\cite{totani}.
A similar mechanism of positron production by collision of 
energetic photons produced in accretion to super--massive central 
black hole and to surrounding primordial black holes with mass 
about $10^{17}$~g was considered in ref.~\cite{titarchuk}.
More exotic scenarios include  annihilating light dark matter 
particles~\cite{silk}, decaying unstable relics and, in particular,
sterile neutrinos~\cite{Picciotto:2004rp},
MeV right-handed neutrino interacting 
with baryonic matter~\cite{Frere:2006hp}, 
strangelets~\cite{Oaknin:2004mn}, positrons originating from primordial
antimatter~\cite{cb-ad}, from decays of milli--charged 
particles~\cite{huh}, and possibly a few more. However, type 
Ia supernovae have a different galactic distribution and their rate is 
probably an order of magnitude smaller than the one necessary to explain
the observed flux~\cite{spi-1, prantzos} (see however~\cite{higdon}). 
Low mass X--ray binaries also do not fit the enhanced 
Bulge component~\cite{spi-1}. Lastly, light dark matter particles
are theoretically not well motivated, might be inconsistent with 
observations~\cite{anti-silk}, and their mass should be very close 
to the electron one, because from the comparison of the Galactic 
gamma ray emission above and below 511 keV, one can conclude that 
the injected energy of the positrons cannot be larger than about 
3~MeV~\cite{beacom}.

In this letter, we discuss the possibility that positrons are 
produced by evaporating primordial Black Holes (BHs). As in the 
case of MeV dark matter, some fine tuning is needed 
too, namely, the BH mass distribution should be peaked in the 
interval $10^{16}-10^{17}$~g. The idea that the primary source 
of low energy positrons in the Galaxy could be primordial BHs 
was first considered in ref.~\cite{okeke}. The picture was later
discussed in ref.~\cite{macgibbon}, with the conclusion that 
primordial BHs could unlikely be responsible for the 511 keV line
from the Galactic Center, unless the primordial BHs are more strongly
clustered in the halo than the other halo material.
However, the authors of ref.~\cite{macgibbon} assumed that the 
initial mass distribution of BHs, $dN/dM$, is scale invariant,
while we have considered a mass distribution which is peaked 
at a particular value. In this case primordial BHs can 
account for the observed 511 keV photons and, at the same time, 
could make even the whole cosmological dark matter, with no 
contradiction with the observed gamma ray backgrounds and 
gravitational lensing data, if the mass spectrum of BHs has a 
pronounced maximum in the interval $10^{16}-10^{17}$ g. 
A mechanism of creation of primordial BH dark matter with a 
peaked mass distribution was suggested e.g. in ref.~\cite{ad-silk}.

{\sc Primordial BHs: general features ---} It is well known that
at the semiclassical level BHs are no longer one--way membranes,
but emit the Hawking radiation~\cite{hawking}. Restricting to the 
simplest case of a Schwarzschild BH with mass $M$, the emission rate of the 
particles of species $i$ with the energy in the range $(E, E+dE)$, 
orbital momentum $l$, third component of the orbital momentum $m$ 
and polarization $s$ is~{\footnote{Throughout 
the paper we use $\hbar = c = k_B = 1$ units.}}:
\be\label{emission}
\frac{dN_{ilms}}{dt} = 
\frac{\Gamma_{ilms}}{(2 \pi)} \frac{dE}{\exp(E/T) \pm 1} \, ,
\ee 
where $\Gamma_{ilms}$ is the so--called graybody factor 
equal to the absorption probability for an incoming wave with the 
specified quantum numbers, $T$ is the Schwarzschild BH temperature 
\be\label{temperature}
T = \frac{1}{8 \pi G_N M} = 1.06 \, 
\left(\frac{\rm 10^{16} \; g}{M}\right) \; {\rm MeV}
\ee
and the signs $\pm$ are for fermions and bosons respectively. At high
energies ($G_N M E \gg 1$), $\sum \Gamma_{ilms} \propto (ME)^2$,
since the cross section for each kind of particle approaches the
geometrical--optics limit, and BH radiates basically as black body 
of the same temperature.

As one can easily see from eq.~(\ref{temperature}), the temperature 
of an astrophysical BH with the mass of the order of the Solar mass 
$M_\odot \approx 2\cdot 10^{33}$ g, or larger, is very low, below $10^{-7}$ K,
and the corresponding particle emission is completely negligible.
However, lighter BHs could have been produced in the very early 
universe by a lot of possible mechanisms (for a review of production 
mechanisms and observational bounds see e.g.~\cite{carr} 
and references therein) and today evaporate at
a much higher temperatures. From eq.~(\ref{emission}), one can 
compute the total particle emission rate and the total
BH mass loss rate~\cite{page, webber}
\be\label{p-loss}
\frac{dN}{dt} &=& \sum_{ilms} \int 
\frac{\Gamma_{ilms}}{\exp(E/T) \pm 1} \frac{dE}{(2\pi)} = 
\frac{\alpha'}{G_N M} = \nonumber\\
&=& 1.1 \cdot 10^{20} \; 
\left(\frac{\alpha'}{2.6 \cdot 10^{-3}}\right)
\left(\frac{10^{16} \; {\rm g}}{M}\right) 
\; {\rm s^{-1}}, \\
\frac{dM}{dt} &=& - \sum_{ilms} \int
\frac{\Gamma_{ilms}}{\exp(E/T) \pm 1} \frac{E \, dE}{(2\pi)} 
= - \frac{\alpha}{G_N^2 M^2} = \nonumber\\
&=& - 4.8 \cdot 10^{20} \; 
\left(\frac{\alpha}{4.5 \cdot 10^{-4}}\right)
\left(\frac{10^{16} \; {\rm g}}{M}\right)^2 
\; {\rm MeV \, s^{-1}}, \nonumber\\ \label{mass-loss}
\ee
where $\alpha'$ and $\alpha$ are numerical coefficients depending 
on the BH mass $M$ and the particle content of the theory. In the
Standard Model, for $M \gg 10^{17} \; {\rm g}$, one finds 
$\alpha' = 1.6 \cdot 10^{-3}$ and $\alpha = 2.8 \cdot 10^{-4}$, 
while, for $5 \cdot 10^{14} \; {\rm g} \ll M \ll 10^{17} \; {\rm g}$, 
$\alpha' = 2.6 \cdot 10^{-3}$ and 
$\alpha = 4.5 \cdot 10^{-4}$~\footnote{For
$M \gg 10^{17}$ g, neutrinos, photons and gravitons are assumed
massless and all the other particles are neglected. For
$5 \cdot 10^{14} \ll M \ll 10^{17}$ g, electrons and positrons
can also be considered massless. The apparent discrepancy with 
ref.~\cite{page} is just due to the inclusion of $\tau$ neutrinos.
The extension of the flux and power factors to include $\tau$ 
neutrinos is given in ref.~\cite{webber}.}. 
From eq.~(\ref{mass-loss}), we can find the BH lifetime
\be\label{tau}
\tau_{evap} &=& \frac{G_N^2 M^3_i}{3 \alpha_i} = \nonumber\\
&=& 15 \; \left(\frac{4.5 \cdot 10^{-4}}{\alpha_i}\right)
\left(\frac{M_i}{5 \cdot 10^{14} 
\; {\rm g}}\right)^3 \; {\rm Gyr} \, ,
\ee
where $\alpha_i = \alpha(M_i)$ and $M_i$ is the initial BH mass,
since the BH spends most of its life near its original mass.
From eq.~(\ref{tau}) we see that all the primordial BHs with the
original mass lighter than $5\cdot 10^{14}$ g have already evaporated. 

{\sc Primordial BHs in our Galaxy ---} Let us start with a
first estimate, just to show that primordial evaporating
BHs can be viable candidates to explain the observed 511 keV 
line from the Galactic Bulge. For this purpose, we assume
that all the BHs have the same mass $M = 6 \cdot 10^{16}$ g. 
The positron production rate per BH is
\be
\frac{dN_{e^+}}{dt} \approx 0.7 \, 
\frac{\alpha_{e^+}'}{G_N M} \approx 2 \cdot 10^{18} 
\; e^+ \; {\rm s^{-1}} \, ,
\ee
where $\alpha_{e^+}' = 0.49 \cdot 10^{-3}$ and the factor 0.7
is the correction due to the finite value of the positron mass. 
In order to explain the observational data, which suggest that 
the positron production rate is $3 \cdot 10^{43}$ $e^+$~s$^{-1}$ 
inside a spherical region of the radius $r = 500 - 700$ pc, 
we need about $1.5 \cdot 10^{25}$ primordial BHs, whose total 
mass is $\sim 4.5 \cdot 10^8 \; M_{\odot}$. Here positrons are 
injected into the Galaxy non--relativistically, together with 
electrons, photons and neutrinos (and possible other unknown 
light particles).

\begin{figure}[b]
\par
\begin{center}
\includegraphics[width=8cm,angle=0]{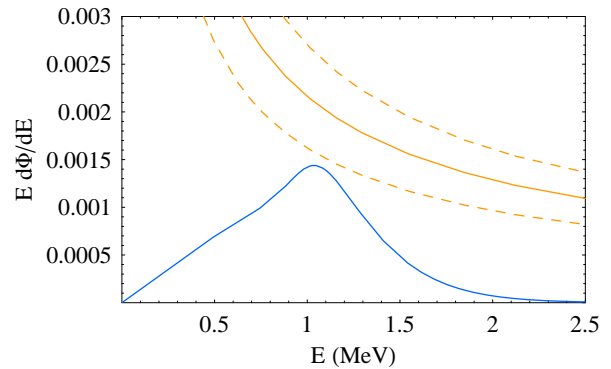}
\end{center}
\par
\vspace{-7mm} 
\caption{Gamma ray spectra from primordial BHs with 
mass $M = 6 \cdot 10^{16}$ g (blue solid curve) and of the diffuse 
background (red dashed curve) in $\gamma \; {\rm cm^{-2} \; s^{-1}}$ 
as a function of energy $E$ in MeV. The number of BHs is 
normalized by the condition that they produce the observed 
positron flux.}
\label{fig}
\end{figure}

The neutrino flux on the Earth would be extremely weak, about
$0.02$ $\nu$ cm$^{-2}$ s$^{-1}$, impossible to be detected by
present and foreseeable future experiments.

The photon flux from the Galactic Center would be about
an order of magnitude less intense than that of neutrinos~\footnote{This 
is due to the fact that neutrino flux receives contributions 
from all three light neutrino generations and that the emission 
rate for spin-1/2 particles is higher than the emission rate for 
spin-1 ones~\cite{page}.}, i.e.
\be
\Phi_{\gamma,{\rm BH}} \approx 1.8 \cdot 10^{-3} \; 
\gamma \; {\rm cm^{-2} \; s^{-1}} \, ,
\ee
and may be observable as a possible bump in the diffuse background 
of the Galactic Center. The presence of the bump depends on the 
exact position of the peak in the photon spectrum from evaporating
BHs. The correct BH photon cross section depends also on the energy of 
the emitted photon, that is $\sigma_\gamma = \sigma_\gamma(M,E)$, 
and has to be computed numerically~\cite{page, webber}. In this
paper we use the photon cross section that one can deduce by 
fitting the curve reported in Fig.~1 of ref.~\cite{webber}. 
The gamma spectrum from primordial BHs is plotted in Fig.~\ref{fig} 
together with the measured Galactic continuum one~\cite{spi-2, continuum}
\be\label{gamma-cont}
\frac{d\Phi_{\rm cont}}{dE} = 
A \left(\frac{E}{0.511 \; {\rm MeV}}\right)^{-1.75} \, ,
\ee
where $A = 7 \cdot 10^{-3}$ $\gamma$ cm$^{-2}$ s$^{-1}$ MeV$^{-1}$
is the normalization factor at $E = 0.511$ MeV.

We can also notice that the cosmological dark matter may be made
of $6 \cdot 10^{16}$~g primordial BHs: their number density today 
would be $n_{BH} = 4 \cdot 10^{-47}$~cm$^{-3}$ and they would have 
created a cosmological MeV gamma background 
$n_\gamma \approx 1.1 \cdot 10^{-11}$~cm$^{-3}$ or, equivalently, an 
isotropic flux $\Phi_\gamma \approx 0.3$~$\gamma$~cm$^{-2}$~s$^{-1}$. 
The latter is about what we observe, so it is not unreasonable that
the Dark Matter (DM) in the Universe is made of BHs with mass 
$M = 6 \cdot 10^{16}$~g.

If the mass fraction of BH with $M = 6 \cdot 10^{16}$ g with respect to
the total amount of DM in the Galactic Center is the same as the 
cosmological average, we would expect about $10^{10} \; M_\odot$ of DM in
the Galactic Bulge. As we see later this is a reasonable result.
On the other hand, we can avoid this restriction if the bulk of DM are
not BHs but is of some other form, e.g. massive stable elementary particles.
In this case the fraction of BHs in the Bulge can be larger than the 
cosmological average and the amount of DM in the Bulge may be 
significantly smaller than $10^{10} \; M_\odot$.

\begin{figure}[b]
\par
\begin{center}
\includegraphics[width=8cm,angle=0]{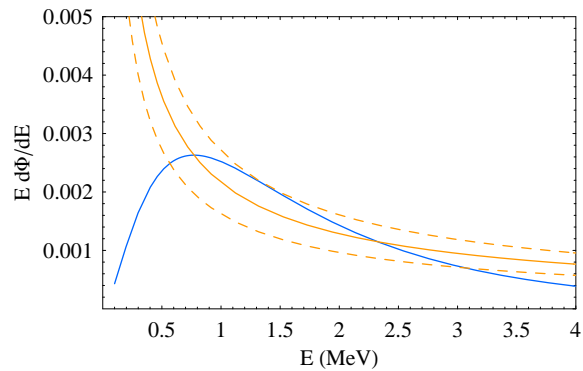}
\end{center}
\par
\vspace{-7mm} 
\caption{Gamma ray spectra from primordial BHs
(dark--blue solid curve) and of the measured background (light--red 
solid curve) from the Galactic Bulge in 
$\gamma \; {\rm cm^{-2} \; s^{-1}}$ as a function of energy 
$E$ in MeV. The BHs are assumed to have log--normal mass 
distribution, eq. (\ref{log-normal}), with the parameters: 
$\gamma = 1$ and $M_0 = 6 \cdot 10^{16}$ g. The number of 
BHs is now normalized by the condition that their total
mass in the innermost 0.6~kpc is $5 \cdot 10^9 \; M_\odot$.
The gamma flux from primordial BHs does not exceed the 
$\pm 25 \%$ uncertainty of the measured gamma ray flux 
(red dashed lines) and can produce enough positrons 
to explain the 511 keV line.}
\label{fig2}
\end{figure}

Keeping an open mind to this possibility, we will concentrate on a more
economical case that all cosmological DM consists of primordial BHs,
considering a more realistic mass distribution than a delta function. 
The mass distribution of such BHs is model dependent and in particular 
strongly depends upon the BH production mechanism 
in the early Universe. One possibility is the model suggested in 
ref.~\cite{ad-silk}, where in the simplest case the BH mass 
distribution has the log--normal form~\footnote{If the density contrast 
appeared as a result of the QCD phase transition, then the masses of 
the BHs created according to the mechanism of ref.~\cite{ad-silk} 
would be cut-off from below by the mass inside the horizon at this 
moment, namely by approximately the solar mass. But the mechanism
allows for generation of large density perturbations at much earlier
stage, e.g. at $T = 10^8 - 10^9$ GeV. In this case the mass inside the
horizon could be as small as $ 10^{16} - 10^{14} $ g.}
\be\label{mass-distr}
\frac{dN}{dM} = C \exp
\left( -\gamma \ln^2 \frac{M}{M_0} \right),
\label{log-normal}
\ee
where $C$, $\gamma$ and $M_0$ are unknown parameters of the
underlying theory. Let us note that this distribution is
rather sharply peaked near mass $M_0$ for $\gamma \geq 1$.
Strictly speaking the distribution given by
eq.~(\ref{mass-distr}) is the initial mass distribution,
and the primordial BH masses changed in the course of cosmological 
evolution. There are two possible effects: evaporation which
leads to mass loss and accretion which results in the
mass increase. The latter leads to mass rise proportional to the
square of the initial mass and weakly change the distribution. The 
evaporation results in complete destruction of primordial BHs with 
small initial masses, $M \leq 5\cdot 10^{14}$~g, and thus to 
decrease the spectrum at small masses. For the taken
here mass distribution with $M_0 \sim 6 \cdot 10^{16}$~g and 
$\gamma \geq 1$ the present mass distribution is quite close 
to the primordial one, because the mass fraction of BHs which 
have already evaporated is small (see below for more details).

\begin{figure}[t]
\par
\begin{center}
\includegraphics[width=8cm,angle=0]{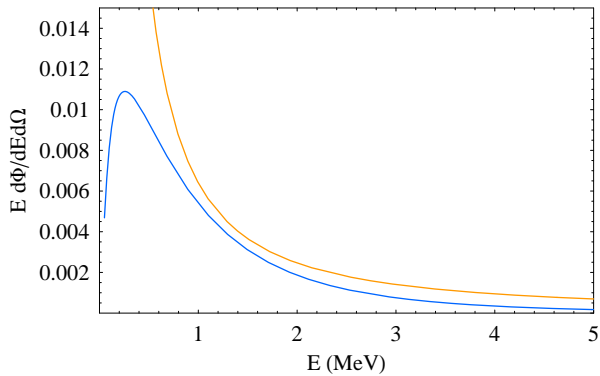}
\end{center}
\par
\vspace{-7mm} 
\caption{Cosmic isotropic gamma ray spectra from primordial BHs
(dark--blue curve) and of the measured background (light--red curve) in 
$\gamma \; {\rm cm^{-2} \; s^{-1} \; sr^{-1}}$ as a function 
of energy $E$ in MeV. Here we assume that the BHs make the whole 
cosmological dark matter and have log--normal mass distribution, 
eq.~(\ref{log-normal}), with the parameters: $\gamma = 1$ and 
$M_0 = 6 \cdot 10^{16}$~g.}
\label{fig3}
\end{figure}

The total amount of DM in the Galactic Center is poorly known. 
The observed rotation curve requires that the total mass in 
the Galactic Bulge is roughly $10^{10} \; M_{\odot}$ and though 
this can be explained by the known baryonic components, see e.g. 
ref.~\cite{bound-dm}, it is natural to expect that the amount 
of DM is of the same order of magnitude as that of baryons.
N--body simulations predict some cusp of DM in the 
center of galaxies~\cite{navarro}, since the typical DM
density profile is $\rho \sim r^{-\beta}$ with $\beta = 0 - 2$. 
In conclusion, $\sim 10^{10} \; M_\odot$ is a reasonable upper 
bound on the amount of DM in the region where most positrons 
annihilate. So, if we assume for example that the total mass 
in the form of primordial BHs in this region is 
$5 \cdot 10^9 \; M_\odot$, we find that the choice $\gamma = 1$ and 
$M_0 = 6 \cdot 10^{16}$~g can explain the 511~keV line flux from 
the Galactic Center and be consistent with the observed galactic 
and extra--galactic gamma backgrounds. This can be seen as follows.
By integrating over all the masses, we find the normalization
constant $C$
\be
\int \frac{dN}{dM} \, dM \Big|_{\gamma = 1, \; 
M_0 = 6 \cdot 10^{16} \; {\rm g}} = 5 \cdot 10^9 \; M_\odot 
\ee
and then we can obtain the photon flux arriving on the Earth
\be
\frac{d \Phi_\gamma}{dE}
= \frac{1}{4 \pi r^2} \int \frac{dN_\gamma}{dE \, dt} \,
\frac{dN}{dM} \, dM \, .
\ee
Here we can use the initial mass distribution, eq.~(\ref{mass-distr}),
and integrate out from 0 to infinity because the corrections due to
BH evaporation are negligible. $dN_\gamma/dEdt$ is provided by 
eq.~(\ref{p-loss}). Fig.~\ref{fig2} shows the gamma spectra from the 
Galactic Bulge produced by the primordial BHs (dark--blue solid curve) 
and the observed gamma background of eq.~(\ref{gamma-cont}) (red solid 
curve). The $\pm$25\% uncertainty of the gamma background
(light--red dashed curves) is at least a reasonable estimate
of the $2\sigma$ curve, see refs.~\cite{spi-2, continuum}.

\begin{figure}[b]
\par
\begin{center}
\includegraphics[width=8cm,angle=0]{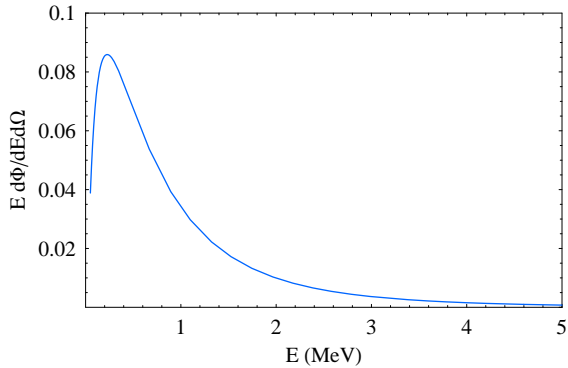}
\end{center}
\par
\vspace{-7mm} 
\caption{Diffuse cosmic neutrino spectrum from primordial BHs
in $\nu \; {\rm cm^{-2} \; s^{-1} \; sr^{-1}}$ as a function of 
energy $E$ in MeV. The assumptions are the same as of Fig.~\ref{fig3}.}
\label{fig4}
\end{figure}

The isotropic cosmological gamma background produced by
primordial BHs in the whole history of the Universe depends
only on the parameter $\gamma$ and $M_0$, if the normalization
constant $C$ is found from the requirement that the primordial
BHs make the whole cosmological DM, whose energy density 
is known to be $\rho_{DM} \approx 2.4 \cdot 10^{-30}$~g~cm$^{-3}$.
The estimate of the diffuse photon flux is (see e.g. 
ref.~\cite{ullio} for the derivation of the formula) 
\be
\frac{d\Phi_{\rm cosmic}}{dE} =
\frac{1}{4\pi} \int \frac{dn}{dM} (M) \, 
\frac{dN_\gamma}{dt dE'}(M,E(1+z)) \,
\frac{dz}{H_0 \, h(z)} \nonumber\\
\ee
where $dn/dM$ is the comoving number density of primordial
BHs with mass $M$, $h(z)=\sqrt{\Omega_{m}(1+z)^3 + \Omega_\Lambda}$
and $H_0$ the Hubble parameter today. The
spectrum is presented in Fig.~\ref{fig3}, together with the 
measured extra--galactic continuum~\cite{extra-continuum}
\be\label{gamma-extra}
\frac{d\Phi_{\rm extra}}{dE} = B 
\left(\frac{E}{1 \; {\rm MeV}}\right)^{-2.38} \, ,
\ee
where $B = 6.4 \cdot 10^{-3}$ $\gamma$~cm$^{-2}$~s$^{-1}$~sr$^{-1}$~MeV$^{-1}$
is the normalization factor. Eq.~(\ref{gamma-extra}) fits quite
well the observed spectrum in the energy range $0.1 - 10$~MeV. 
Contrary to the simplest case of primordial BHs with equal masses, 
i.e. with the delta-function spectrum, primordial BHs with sufficiently 
wide spectrum may explain the observed annihilation line, be
in agreement with the observed gamma background, and make  
all cosmological DM.

\begin{figure}[t]
\par
\begin{center}
\includegraphics[width=8cm,angle=0]{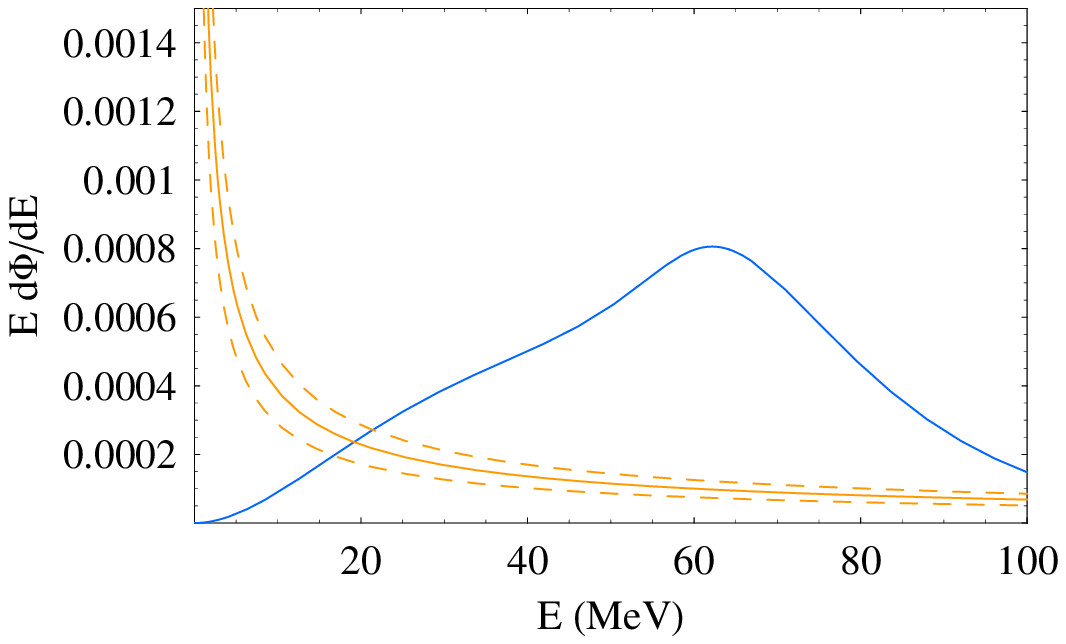} \\
\includegraphics[width=8cm,angle=0]{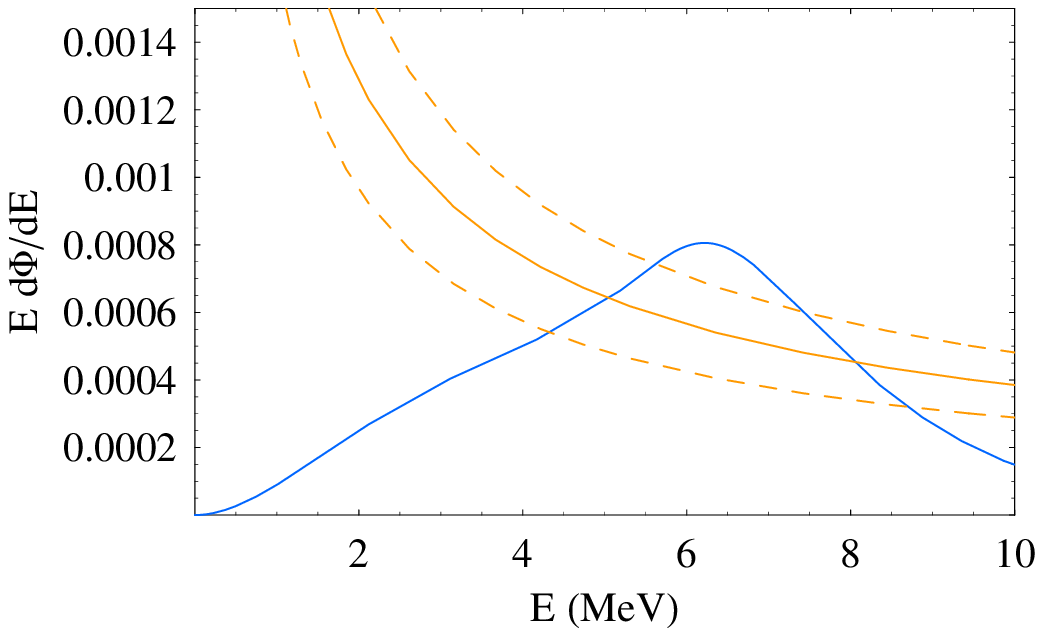} \\
\includegraphics[width=8cm,angle=0]{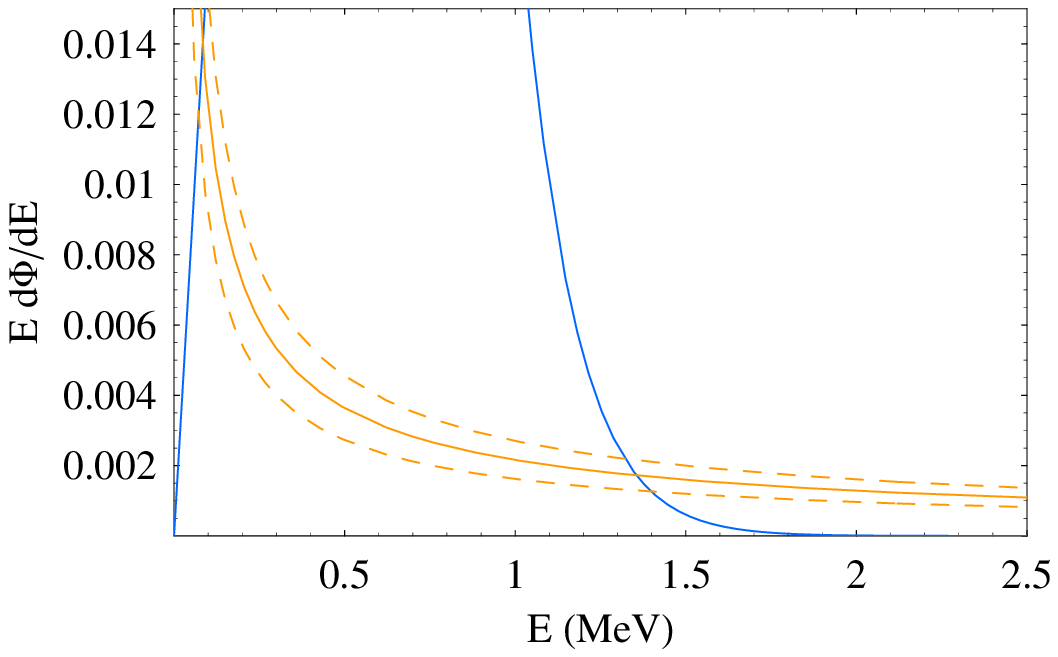}
\end{center}
\par
\vspace{-7mm} 
\caption{Gamma ray spectra from primordial BHs (dark--blue solid curve) 
and of the measured background (light--red solid curve) from the Galactic 
Bulge in $\gamma \; {\rm cm^{-2} \; s^{-1}}$ as a function of 
energy $E$ in MeV. The BHs are assumed to have all the same mass:
$M = 10^{15}$~g (upper panel), $M = 10^{16}$~g (central 
panel) and $M = 10^{17}$~g (lower panel). The number of BHs is 
normalized by the condition that they produce the right amount
of positrons to explain the observed 511~keV line. Red dashed lines
are the $\pm$25\% uncertainty of the measured diffuse gamma flux.}
\label{fig5}
\end{figure}

Fig.~\ref{fig4} shows the expected cosmological neutrino background
from the evaporating primordial BHs. Unfortunately, such neutrinos
are difficult to observe, because they are below the solar and 
terrestrial neutrino fluxes.

One could also wonder about the impact of the primordial BH 
positron and electron emission spectra on the diffuse Galactic 
backgrounds. Simple considerations suggest that MeV positrons 
and electrons produced in the Galactic Bulge are confined to 
the Galactic Bulge~\cite{silk} and therefore their flux on
the Earth is negligible. Even if we take into account the fact 
that such primordial BHs could make the whole cosmological 
DM, and thus some of them are also in the neighborhood of 
the Solar System, one cannot expect any effect on the measured
Galactic background, because most of the positrons are 
non--relativistic and are likely unable to reach a detector 
which is located on a satellite orbiting around the Earth.

The positrons produced by primordial BHs have low energies
and there is no contradiction with the bound deduced in 
ref.~\cite{beacom}. Actually, the constraint of 3~MeV is 
obtained assuming that all the positrons have the same energy, 
so it is not easy to apply the result to more general cases.
However, we can quickly see that in the scenario proposed in this 
paper the positrons with energies above 3~MeV are about 5\%
of the total number of positrons produced by Hawking evaporation.
This is found calculating the integral   
\be
\int \, dM \, M^2 \frac{dN}{dM} \,  
\int_{3 \; {\rm MeV}}^{\infty} \, dE \,
\frac{E \sqrt{E^2 - m_e^2}}{1 + \exp(E/T)}
\ee
and dividing by the same integral evaluated between $m_e$ and 
infinity. Including the grey--body effect, the correction to
this estimate is irrelevant, close to a factor 1.1.

The mass distribution of eq.~(\ref{mass-distr}) includes also
very light BHs which have already evaporated. For our
choice $\gamma = 1$ and $M_0 = 6 \cdot 10^{16}$~g, the initial
fraction of BHs with masses smaller than $5 \cdot 10^{14}$~g,
and hence lifetimes shorter than the present age of the Universe,
is negligible, at the level of $10^{-16}$. However, the total
amount of matter converted into relativistic particles from the 
birth of the Universe up to today is much larger: the dominant
contribution comes from BHs which still exist in the Universe
and are basically evaporating at a constant rate, because the
process is slow and their mass is still very close to the 
initial one. Assuming that primordial BHs make the all DM,
today the cosmic energy density of the particles produced
by BH evaporation is roughly $10^{-7} \; \rho_{DM}$. 
The photon flux is the one reported in Fig.~\ref{fig3}.

Lastly, we would like to note that the BH masses are strongly
constrained by the gamma ray continuum from the Galactic Center.
Indeed, neglecting any possible correlation between primordial BHs 
and DM, we can consider three simple cases, where all the BHs have 
the same mass: $M = 10^{15}$~g, $M = 10^{16}$~g and 
$M = 10^{17}$~g. In order to explain the observed 511 keV gamma flux, 
the total mass of primordial BHs in the Galactic Bulge should 
respectively be about $7 \cdot 10^4 \; M_\odot$, 
$7 \cdot 10^6 \; M_\odot$ and $9 \cdot 10^{10} \; M_\odot$. The 
produced galactic photon background for these cases is presented in 
Fig.~\ref{fig5}. For low BH masses, the BH temperature is high and too 
many high energy photons are produced. For high masses, the positron 
emission is exponentially suppressed by the Boltzman factor and to
explain the observed 511~keV line we would need a large number of BHs, 
with the effect that the low energy gamma spectrum is too intense. 
Hence, we can conclude that, assuming that the positron flux is produced 
by primordial BHs, the BH mass distribution has to be peaked around 
$10^6$~g and actually rather sharply peaked near $6 \cdot 10^{16}$~g.

{\sc Conclusion ---} We have considered evaporating primordial BHs,
as a possible source of positrons to generate the observed photon
511 keV line from the Galactic Bulge. The analysis of the accompanying 
continuous photon background produced, in particular, by the same 
evaporating BHs, allows to fix the mass of the evaporating BHs 
near $10^{16}$ g. It is interesting that the 
necessary amount of BHs could be of the same order of 
magnitude as the amount of dark matter in the Galactic 
Bulge. This opens a possibility that such primordial BHs may 
form all cosmological dark matter. The 
background MeV photons 
created by these primordial BHs can be registered in the near future,
while the neutrino flux may be still beyond observation.
The significance of this model 
would be difficult to overestimate, because these BHs would present 
a unique link connecting early universe and particle physics.

After this paper was completed, we became aware of the 
paper~\cite{Frampton:2005fk}
where a similar idea was explored. However, the authors of this 
work assumed much lighter BHs
which is in contradiction with our results.

\begin{acknowledgments}
C.B. and A.A.P. are supported in part by NSF under grant PHY-0547794 
and by DOE under contract DE-FG02-96ER41005.
\end{acknowledgments}


\newpage

\section*{Erratum}

We overestimated the total annihilation rate in the 
Galaxy almost by factor 2. Instead of the value
\be
3 \cdot 10^{43} \; {\rm s^{-1}} \, ,
\ee
used in the paper, the correct one is 
\be
1.6 \cdot 10^{43} \; {\rm s^{-1}} \, .
\ee 
Our error was based on the statement of ref.~[6] of 
our paper where it is written that ``the true 
annihilation rate is 3.6 times larger than would be 
deduced from the 0.511~MeV flux alone''. The corrected 
number of the annihilation rate results in shift 
down of the theoretical curves in figs.~1 and 5 by 
factor 1.8 -- 1.9. We thank Pierre Jean for indicating 
to this error.

\end{document}